\documentclass[a4paper,11pt]{article}
\pdfoutput=1 
\usepackage{jinstpub} 
\usepackage{lineno}

\newcommand{\comment}[1]{}

\title{Gadolinium Loaded Cherenkov Detectors for Neutron Monitoring in High Energy Air Showers  }

\author[a]{P. Stowell \note{Corresponding author.}}
\author[b]{S. Fargher}
\author[b]{L. F. Thompson}
\author[a]{A. M. Brown}
\author[a]{P. M. Chadwick}

\affiliation[a]{University of Durham, UK}
\affiliation[b]{University of Sheffield, UK}
\emailAdd{john.p.stowell@durham.ac.uk}


\abstract{Monitoring of high energy cosmic ray neutrons is of particular interest for cosmic ray water Cherenkov detectors as intense bundles of delayed neutrons have been found to arrive after the initial passage of a high energy air shower. In this paper we explore the possibility of building large-area high-energy neutron monitors using gadolinium-loaded Water Cherenkov Detectors (WCDs). GEANT4 simulations of photon production in WCDs are used to estimate the maximum detection efficiency for a hypothetical system. Requiring a series of neutron induced gamma ray flashes distributed over an extended period of time (up to 20$\mu$s) was shown to be an effective way to discriminate high energy neutron interactions from other backgrounds. Results suggest that neutron detection efficiencies of 4-15\% may be possible using a gadolinium-loaded detection system above 200~MeV. The magnitude of gadolinium loading was also shown to significantly modify the timing response of the simulated detector.}

\begin{document}
\maketitle
\flushbottom

\section{Introduction}
\label{sec:introduction}

Measurement of delayed low-energy hadrons after the arrival of a cosmic-ray Extended Air Shower (EAS) can be an effective probe to determine whether an EAS is hadronic in origin. Neutrons have been found to arrive up to a millisecond after the detection of the core electromagnetic component of an EAS \cite{balabin2011eas}. Simulation studies have reported that the delay is largely dependent on the neutron energy, with a flux of much lower energy thermalised neutrons arriving many milliseconds after the initial EAS \cite{engel2021neutron}. Understanding the temporal and spatial extent of this delayed neutron component in EAS is of particular interest for water Cherenkov and scintillator based high energy cosmic ray observatories as they can be sensitive to secondary neutrons interacting in detection systems, and some studies have found that there is a turn-on in delayed neutron generation at energies close to the knee-region in the cosmic ray spectrum \cite{antonova2002anomalous,shepetov2020measurements}.

Driven by several observations of this sub-luminal neutron `cloud'  \cite{linsley1984sub,stenkin2007neutrons}, and reduction in cost of thermal neutron detectors, significant progress has been made in large area monitoring of the low energy neutron component \cite{levochkinelectron}. Recently, the LHAASO experiment has published data on the time evolution of the thermal neutron component following an EAS using a distributed array of boron based thermal neutron detectors \cite{levochkin2021study}.  Given the complex and stochastic scattering processes that neutrons undergo as they thermalise with their surroundings, simultaneous measurements of both low and high energy neutrons across the full extent of an EAS are needed to understand the evolution of the neutron `cloud' and its effects on water Cherenkov based cosmic ray detector arrays. In particular it has been shown that some component of the neutron signals observed in an extended air shower are the result of high energy electromagnetic particles interacting with detector materials, amplifying the delayed neutron signal \cite{erlykin2007neutron}.

Typically studies investigating the arrival time and total energy contribution of the high energy neutron component of EAS have deployed a single or small number of NM64 neutron detectors  \cite{hatton1964experimental} to try to measure neutron fluxes close to the EAS core \cite{badalyan2017extensive}. These systems, based on boron-triflouride proportional tubes surrounded in lead, are the adopted standard for high energy cosmic-ray neutron detection, and a number of them are in use today as part of a worldwide cosmic ray monitoring network. This `neutron monitoring database' provides continuous measurements of the temporal variation in cosmic ray intensity using an array of neutron detectors distributed across the globe at varying geomagnetic cut-off rigidities (500 MeV to 15 GeV). 
The NM64 design (based on an earlier design by Simpson \cite{simpson1957cosmic}) is optimised for the detection of high energy neutrons by taking advantage of the fact that fast neutrons produce a shower of secondary lower energy evaporation neutrons when striking a high mass target. The NM64 stations in the global neutron monitoring network have been shown to be sensitive to rapid increases in the ground level cosmic ray intensity as a result of intense coronal mass ejections \cite{mishev2014analysis}. They therefore provide a useful alarm system for ground level enhancement (GLE) events that pose a risk to aircraft or on-ground infrastructure. However, the rapid identification of GLE's to issue an effective global warning  requires large area detectors with a high enough counting rate so that increases to the cosmic ray intensity can be determined with a high statistical significance on the order of several minutes. New techniques for producing large-area high-energy neutron detectors could be used to further extend the neutron monitoring database and reduce the time required to issue GLE warnings on a global scale.

Recently, several neutrino experiments have started to investigate the possibility of using gadolinium loaded water Cherenkov detectors for efficient large volume neutron identification \cite{xu2016current,vagins2019gadolinium,laha2014gadolinium}. Gadolinium has a high thermal neutron absorption cross-section, and produces two gamma rays with combined energy of up to 8 MeV following a neutron capture. Since gadolinium sulphate can be readily dissolved into water, the loading of gadolinium into water Cherenkov tanks has been shown to be an effective way to build a large volume neutron detector. Furthermore, since water acts as both a passive neutron moderator and an active gamma ray detection medium, the combination results in a high efficiency neutron detector using  relatively small quantities of gadolinium. The Super-Kamiokande water Cherenkov experiment is currently exploring gadolinium loading \cite{abe2021first}, and initial tests found a loading of  0.2\% by weight can achieve a neutron trigger efficiency of approximately 66\% in inverse beta decay reactions \cite{watanabe2009first}. 

There are a number of similarities in detector designs and readout methodologies for these experiments and future water Cherenkov air shower experiments. It is therefore interesting to explore whether Gd-loaded water tanks could also be used to identify high energy neutrons in EAS in an effort to reduce air shower reconstruction uncertainties. Gd-loaded WCDs are expected to be primarily sensitive to thermal neutrons in an EAS, however the addition of high atomic mass materials, such as lead blocks, is expected to increase the likelihood of neutron spallation events occuring in the detectors, making them sensitive to higher energy muons or hadrons.  This is if particular interest for high altitude gamma ray observatories such as HAWC \cite{abeysekara2017observation} or the future Southern Wide-field Gamma-ray Observatory (SWGO) \cite{abreu2019southern}. In HAWC EAS originating from high energy gamma rays must be discriminated from a significantly larger galactic cosmic ray EAS background which are hadronic in origin \cite{hampel2015gamma}. This is achieved by assessing the compactness and relative smoothness of the EAS, since hadron induced showers typically produce clustered high energy deposition regions far from the shower core. This background rejection methodology, whilst effective, has a degraded performance when the core falls close to the edge of the detector array or photon statistics limits the ability to clearly identify clusters  away from the core. Since the direct measurement of even a single high energy muon or hadron at ground level is a strong indication an EAS is hadronic in origin, new detectors capable of measuring the hadronic contributions across multiple energy scales in an EAS  could significantly improve background rejection in gamma ray observatory detector arrays.

This paper uses GEANT4 \cite{agostinelli2003geant4} to assess the feasibility of constructing high energy neutron monitors using lead-lined gadolinium loaded water Cherenkov tanks. Section~\ref{sec:detgeo} describes the detector principle and an assumed detector geometry considered in this simulation work, Section~\ref{sec:startgeo} describes the GEANT4 simulations of a hypothetical detector for a fixed gadolinium loading, Section~\ref{sec:gadvar} discusses the effect different amounts of gadolinium loading could have on particle discrimination, and finally Section~\ref{sec:conc} discusses the  potential application of this system for long term cosmic ray monitoring and hadronic content quantification in extended air showers.

\section{Detector Geometry}
\label{sec:detgeo}

One problem with the use of Gadolinium (Gd) as a neutron capture agent is that large quantities of moderator material are needed for a detector to have a good efficiency when detecting the highest energy neutrons present in cosmic rays. This problem is common not only for gadolinium, but also helium-3, and boron-triflouride systems which have cross-sections which are at a maximum below 1~eV. The NM64 neutron detector design uses additional lead `producer' materials to overcome this challenge when targeting only the high energy region of the neutron spectrum. The probability of neutron-induced spallation producing showers of secondary hadrons increases with both neutron energy and target atomic mass. Interaction of a high energy neutron with a lead `producer' target typically produces several secondary lower energy evaporation neutrons, shifting the neutron energy spectrum to a region with higher capture efficiency. In addition this helps to increase the signal-to-noise ratio of a detector as it is likely multiple neutrons are detected within the integration window of neutron sensitive detector placed next to the lead producer.

\begin{figure}
    \centering
    \includegraphics[width=\textwidth, clip, trim=0 10 0 0]{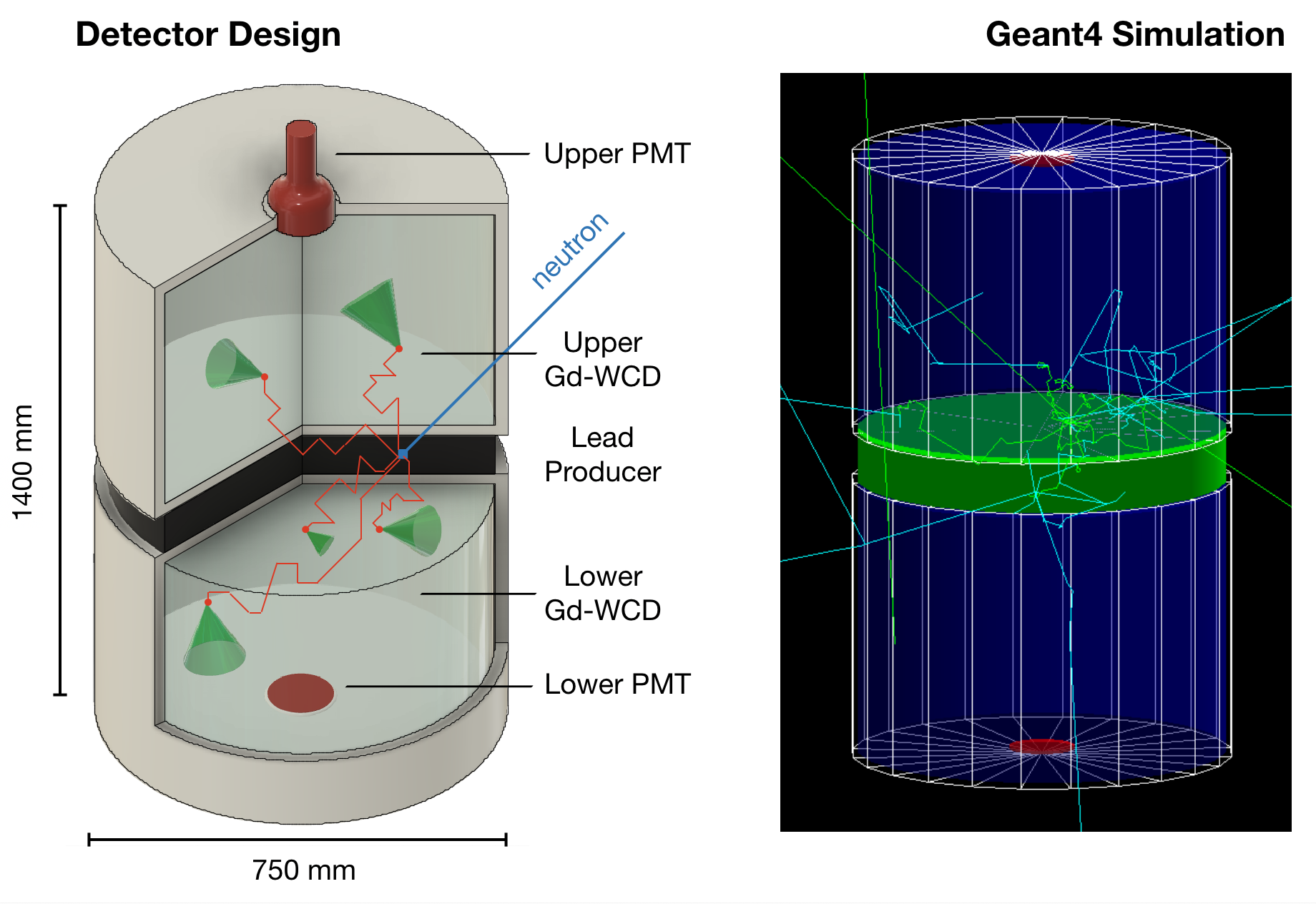}
    \caption{Detector design concept. High energy neutrons undergo spallation inside the lead producer, resulting in a shower of hadrons diffusing outwards from the primary interaction point. Neutrons in this shower are moderated and captured on gadolinium in the water to produce a train of distinct gamma ray Cherenkov `flashes'. (Left) Simulated tank dimensions. (Right) GEANT4 simulation showing an example spallation event.}
    \label{fig:tankdesign}
\end{figure}

In this work we propose a novel combination of the NM64 `producer and detector' design and the recent work on Gd-loaded Water Cherenkov Detectors (Gd-WCDs). As shown in Figure \ref{fig:tankdesign}, a lead producer in the centre of the detector is surrounded by two Cherenkov tanks filled with water loaded with a small concentration of gadolinium sulphate. Instead of additional layers of moderating HDPE as in the NM64 design, the water itself acts as an active moderator layer around the lead producer target, maximising spallation neutron detection efficiency. For high energy neutron spallation events it is likely that the secondary neutrons produced undergo a random walk in all directions from the interaction point. Requiring hits in both the upper and lower water Cherenkov detectors helps to reject backgrounds from low energy electron and gamma-induced interactions in one tank.
The only particles expected to produce a dual tank coincidence are high energy neutrons, muons, and a small proportion of high energy electromagnetic showers that are not ranged out by the lead producer acting as a shield between the two tanks.  Furthermore, due to the stochastic nature of low energy neutron propagation through the tank, a neutron spallation event is expected to produce a clear timing signature of several gamma ray flashes over approximately 100~$\mu$s, which can be recognized by digital pulse shape processing.

In this  work we consider a hypothetical design to demonstrate the feasibility of this detection principle when considering the constant flux of background cosmic rays at sea level. The hypothetical detector is assumed to be comprised of two cylindrical 650~mm x 750~mm water Cherenkov tanks, placed above and below a lead producer target of 10~cm thickness. The tank is assumed to be filled with water and a variable gadolinium-sulphate loading concentration. The choice of materials, and tank size have been chosen at this stage to support future work to develop and build a small scale prototype. Exact dimensions are arbitrary however, as the present work serves only to demonstrate the feasibility of using such a system to monitor the high energy hadronic component of cosmic ray air showers.

\section{GEANT4 Simulations}
\label{sec:startgeo}

GEANT4 simulations of Cherenkov light production within the hypothetical detector have been used to assess particle discrimination performance \cite{agostinelli2003geant4}. Propagation of particles within GEANT4 is handled by the QGSP\_BERT\_HP physics list with an extension to include production of Cherenkov photons within water. As shown in Figure \ref{fig:tankdesign}, the detector itself is treated as two uniform cylinders of water placed either side of the lead producer. The water is assumed to have a density of 1~g cm$^{-3}$, and a simplified assumption of a constant refractive index of 1.333 across all wavelengths. For a real detector the choice of tank reflector materials and photomultiplier coverage all factor in to the overall neutron detection efficiency. 

This represents a large parameter space that needs to be optimised, therefore to avoid numerous biases at this early stage, we count the total number of Cherenkov photons produced at each optical photon's production point instead of fully propagating them through the detector and towards the readout PhotoMultiplier Tubes (PMTs). This gives an estimate of the maximum number of Cherenkov photons that may be detectable for a given type of interaction and allows us to assess the underlying feasibility of the detection principle before additional light collection systematic uncertainties are included. The ground is assumed to be a 30~m $\times$ 30~m area, 3~m deep box of dry calciferous rock directly below the detector.

For a realistic estimation of the particle distribution from cosmic rays, the CRY cosmic ray generator library is interfaced with GEANT4. This generates a flux of primary particles from a 30~m $\times$ 30~m source plane 2~m above the detector, and automatically updates the simulations expected `real exposure time' so that overall trigger rates can be estimated. Date, latitude and altitude are required for accurate CRY simulations. The date was  set to  2021-01-01, and the latitude and altitude were chosen to be sea level at 54$^\circ$7671~N (Durham, UK Latitude). This was chosen to provide an early indicator of counting rates to support future work assembling a small scale prototype of this system. The CRY library provides estimates of the relative proportion of different cosmic ray particle species for the ever-present cosmic ray flux at the surface. Based on a given source plane area it estimates the likelihood of two or more cosmic ray particles being generated at a given time and so provides a simplistic way to consider the effects of event pile-up on chosen trigger conditions. However, given their rarity, the CRY simulation does not include statistically significant simulations of very high energy distributed air showers. The results shown here therefore consider only whether the hypothetical detector system can reliably distinguish high energy neutrons (>1 MeV) from lower energy cosmic rays. Understanding detection efficiency and pile-up effects in ultra high energy extended air showers is the topic of future work.

\begin{figure}
    \centering
    \includegraphics[width=1.0\textwidth, clip, trim=0 10 0 0]{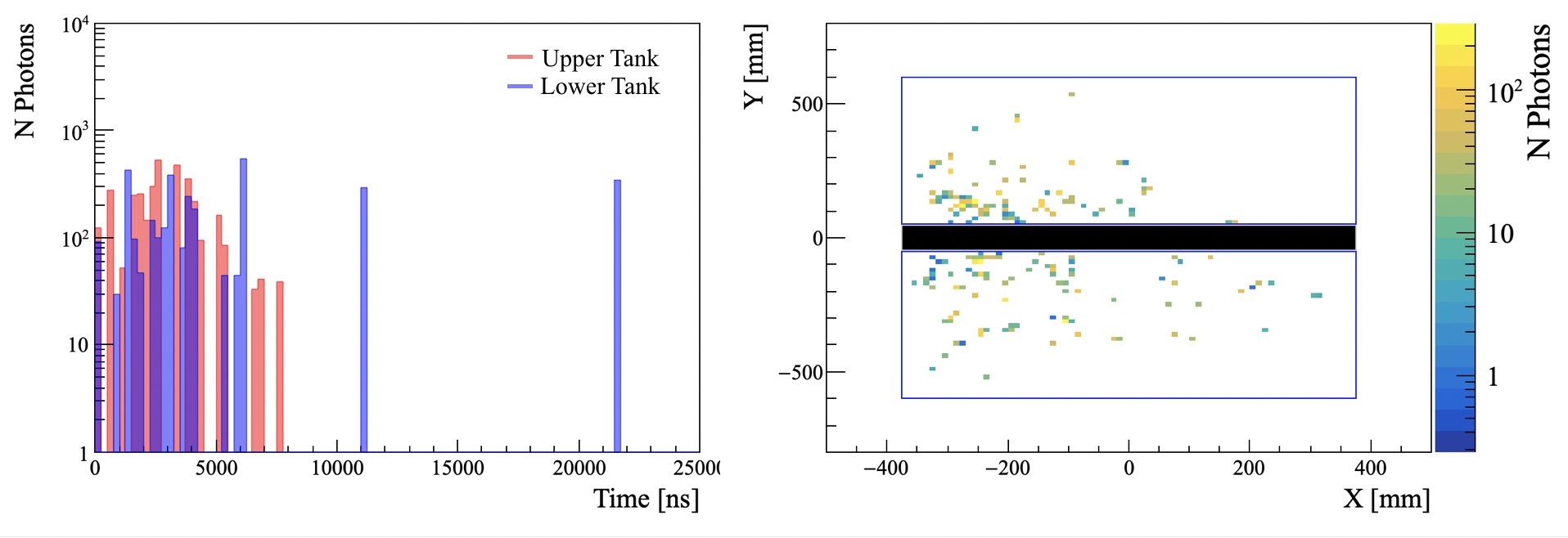}
    \caption{Example Neutron Spallation Event. (Left) The number of true Cherenkov photons produced in each timing bin for both the upper and lower tank. (Right) Distribution of Cherenkov photon production points as a function of position in both tanks.}
    \label{fig:tankpulse}
\end{figure}

To assess a suitable trigger condition, a baseline design consisting of two Gd-WCDs with an assumed Gd loading of 0.2\% by weight was first evaluated. 
A Monte-Carlo simulated event sample was generated by requesting approximately 3600~s of exposure time from the CRY-GEANT4 interface.  
For each simulated event, primary particle information and the production point and timestamp of every Cherenkov photon generated was saved. 
This provides an estimate of the photon timing profile of individual events before effects such as tank internal reflectance and photomultiplier readout are included, making it possible to estimate the feasibility of the proposed detection technique at a fundamental level. 
For each event, a simplified `offline processing' was performed to mimic the digitization of a real event. 
The time of production of the first Cherenkov photon was used as the trigger time, `$t_{0}$', with all other photon timestamps being calculated relative to this time. 
The relative timestamps of all photons produced in a single event were then filled into coarse histogram bins of width 250~ns for both the upper and lower simulated water Cherenkov tanks. 
This provides an estimate of the timing distribution of pulses that might be observed by a photomultiplier tube in a real experiment. 
Figure \ref{fig:tankpulse} shows a timing distribution for a single neutron spallation event, and the corresponding distribution of Cherenkov photon production points within both tanks. 
A neutron spallation event is typically followed by a series of discrete gamma ray `flashes' as the hadronic shower diffuses and thermalises before neutrons captured. 
Whilst the choice of tank materials could modify this timing distribution slightly, given the relatively coarse resolution of the binning considered, these effects are likely to affect only the overall amplitude of photons in each timing bin, not their relative distribution in time. 
Multiple muons arriving within a 250~ns window have been considered as a triggering scheme for the efficient detection of muon bundles in air showers \cite{kokoulin2015seasonal}, therefore analysing the data using this coarse binning scheme provides a triggering method that is unlikely to see large background contributions from a single muon bundle in future work.

\begin{figure}
    \centering
    \includegraphics[width=0.49\textwidth, clip, trim=0 0 0 0]{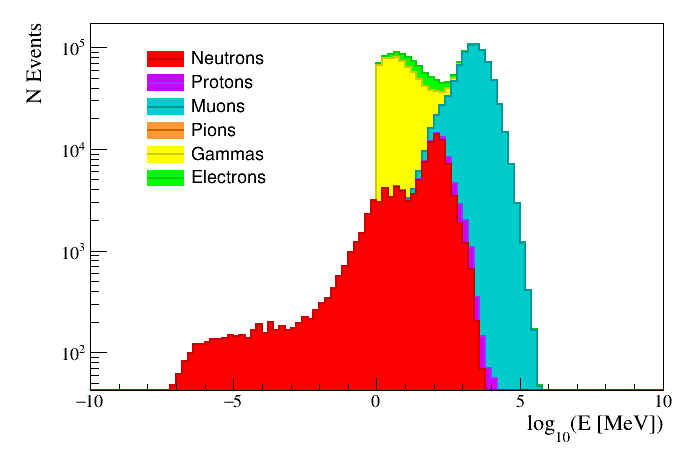}
    \includegraphics[width=0.49\textwidth, clip, trim=0 0 0 0]{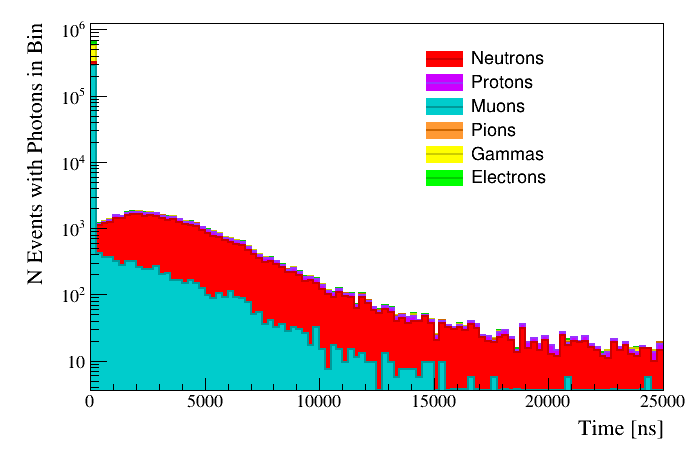}
    \caption{Different particle species distributions before a two-tank coincident cut is applied. (Left) Energy distribution for remaining events separated by particle species. (Right) Timing distribution for all remaining events separated by particle species. Neutron and proton induced spallation events produce a series of neutron capture flashes that follow the first flash observed resulting in a second peak after approximately 3$~\mu$s. }
    \label{fig:prespecies}
\end{figure}

The possibility of event pile-up based on the CRY cosmic ray rate is included by looking forwards in simulated time for each event and including all photons from subsequent timestamps which are within a 250~$\mu$s range. Including this was not found to have a significant effect on the results. The timing distribution histograms for each event can be added together to estimate the average timing distribution for any given starting condition generated by CRY. Figure \ref{fig:prespecies} (left) shows this timing distribution for different cosmic ray particle species.  
Figure \ref{fig:prespecies} (right) shows the energy and timing distributions for cosmic ray particles that produced photons in either the top or bottom tank. The sharp cutoff in the gamma energy distribution below 1~MeV is a known discontinuity in the CRY model due to a minimum energy for these particles. It is clear that whilst a high proportion of interactions comes from gamma and electron cosmic ray events, these events typically occur in an short space of time, with all photons falling almost entirely within the first 250~$\mu$s bin. Lower energy gammas are expected to deposit their energy in a similar way. The only events that produce a wider photon timing spread are therefore a result of high energy neutron, muon, and proton induced spallation events.

\begin{figure}
    \centering
    \includegraphics[width=0.49\textwidth, clip, trim=0 0 0 0]{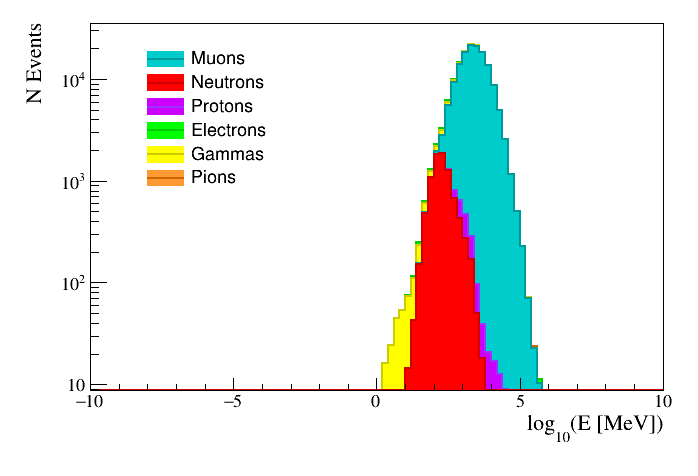}
    \includegraphics[width=0.49\textwidth, clip, trim=0 0 0 0]{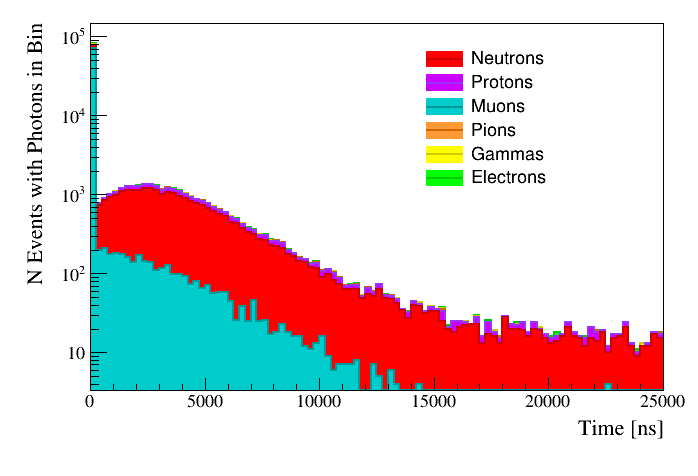}
    \caption{Different particle species distributions after a two-tank coincident cut is applied. (Left) Energy distribution for remaining events separated by particle species. Only the highest energy component of the cosmic ray neutron spectrum produces a two-tank Cherenkov flash signal. (Right) Timing distribution for all remaining events separated by particle species. Neutrons and proton induced spallation events produce a series of neutron capture flashes that follow the first flash observed. }
    \label{fig:species}
\end{figure}

\begin{figure}
    \centering
  \includegraphics[width=0.49\textwidth]{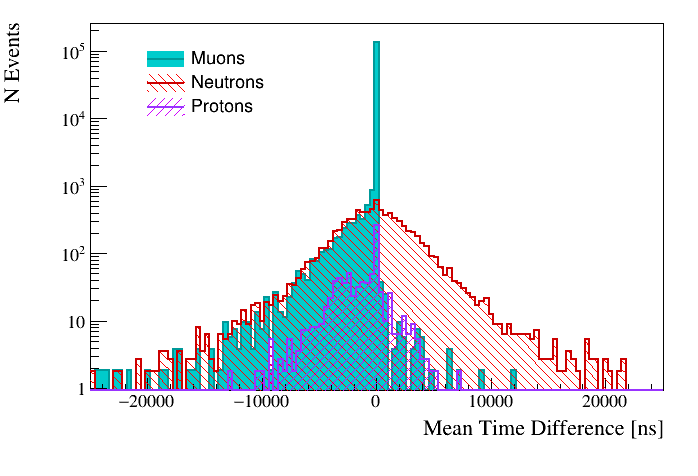}
  \includegraphics[width=0.49\textwidth]{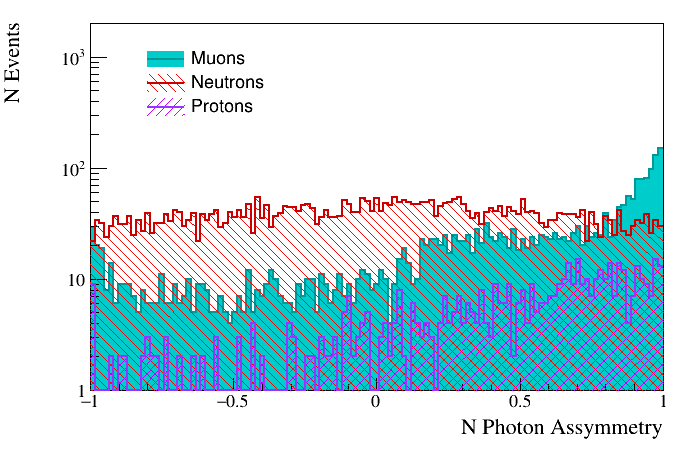}
    \caption{Distribution of time differences and pulse time asymmetry for different generated particle species. (Left) The mean time difference between the top and bottom tank shows a clear bias towards the first photons being produced in the upper tank for charged particles. (Right) The N Photon asymmetry is calculated from the relative difference in the number of photons for both the upper and lower Gd-WCD tanks, showing that muons produce the majority of their photons in the upper tank. }
    \label{fig:assymetry}
\end{figure}

A requirement for photons to be produced in both the top and bottom Cherenkov tanks within this 250~$\mu$s time period was found to reject a significant number of cosmic ray-induced background events caused by electron/gamma induced showers, or single low energy neutrons being captured in only one of the tanks. Additional background radiation from the surrounding soil was neglected in all simulations as the likelihood of coincident light production from a single gamma ray is expected to be low. A high gamma background is capable of reducing the overall signal-to-noise of the proposed detector; however, since this is highly site-dependent, the effect is neglected in the present work. As shown in Figure \ref{fig:species}, after a  coincidence selection criterion has  removed most of the low-energy electromagnetic background, only a combination of through-going muons, gammas, and hadrons remain. It is clear that only the highest energy component of the neutron spectrum results in coincident photon production in both tanks. This technique therefore provides a way to measure the high energy cosmic ray neutron rate without including additional uncertainties from low energy neutrons that may have thermalised in the surrounding soil. From the average timing distributions it is clear that muons still deposit the majority of their energy within the first histogram bin (within 250~ns), whilst the average timing distribution of neutron and proton interactions can be spread over many micro-seconds. For a tank with Gd loading of 0.2\% the majority of the secondary neutron capture flashes occur within a  10-15~$\mu$s window. This is a similar order of magnitude to the typical hold-off time used for a NM64 neutron detector (20~$\mu$s). It is therefore possible to distinguish high energy hadrons from muons with a reasonably high purity based solely on the combined timing distribution of photons produced in both the upper and lower Gd-WCDs. Am asymmetry metric can be calculated based on the number of event photons in each tank  as $(N_{top}-N_{bottom})/(N_{top}+N_{bottom})$ where $N_{top}$ and $N_{bottom}$ are the total number of photons in the top and bottom tank respectively. As shown in Figure \ref{fig:assymetry} either the difference in the average photon production time or the `N Photon Asymmetry` metric can be used to effectively discriminate muon-induced spallation events.  This discrimination is possible because downward going muons produce a much larger number of Cherenkov photons as they pass through the upper Gd-WCD before they induce a secondary set of delayed neutron flashes. 

\begin{figure}
    \centering
 \includegraphics[width=0.49\textwidth, clip, trim=0 0 0 0]{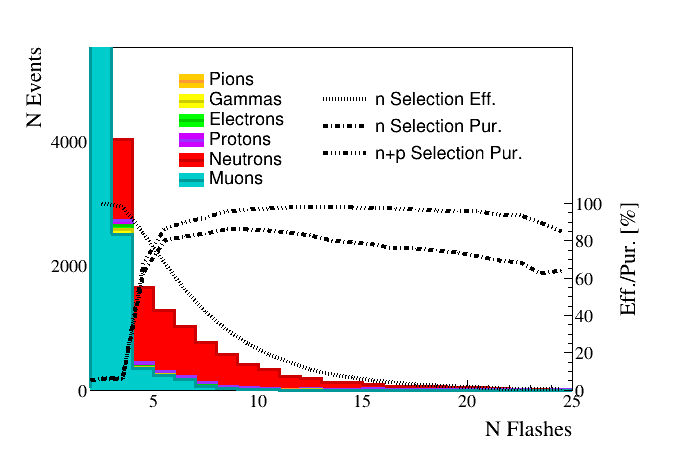}
 \includegraphics[width=0.49\textwidth, clip, trim=0  0 0 0]{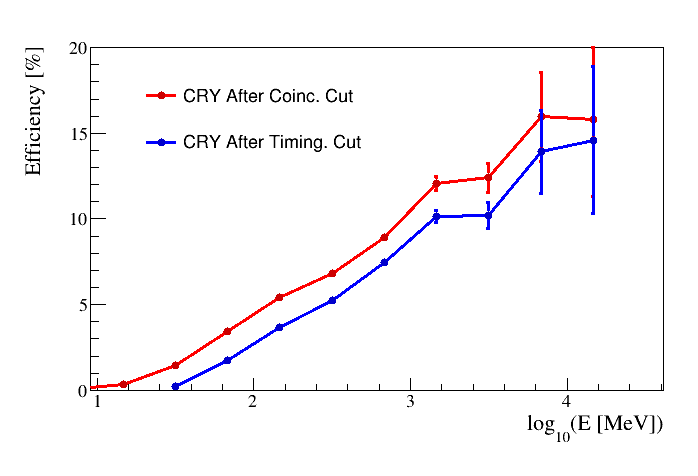}
    \caption{Timing Bin Efficiency Plots. (Left) Number of timing bins with non-zero entries for each event separated by particle species. A cut of at least four bins being non-zero was chosen as a simple trigger approximation. (Right) Neutron selection efficiency before and after the N Flashes cut. }
    \label{fig:bincut}
\end{figure}

To estimate the detection efficiency as a function of generated neutron energy, a simplistic particle discrimination technique is assumed. Based on the number of timing bins with non-zero contributions shown in Figure~\ref{fig:bincut}, an event is considered to be a neutron when the total bins filled in both upper and lower tanks is greater than four. Given their relatively small contribution, no additional cuts to remove muon-induced spallation events were considered at this stage. For the sample of events shown previously this results in an overall neutron selection efficiency of 2.48\% and a purity of 71.3\% for neutrons (89.2\% for neutrons and protons). This efficiency is based on the starting number of neutrons over all simulated events, including lower energy neutrons which scattered before entering the detector itself. When plotted against energy, the efficiency peaks above 10~MeV for Gd-WCDs, with an average efficiency of 4.58\% between $10^2$~MeV and $10^4$~MeV. The simplistic triggering condition explored here is therefore a viable technique for the detection of the highest energy component of neutrons in cosmic ray air showers. It is possible that a deeper analysis of triggering optimisation using finer bins and an in-depth photon propagation simulation could obtain improved detection efficiencies, however as shown in Figure \ref{fig:bincut} the total possible improvement is marginal when considering the overall efficiency between $10^2$~MeV and $10^4$~MeV. It is therefore more likely that modifications to the lead producer configuration or Gd-WCD are needed to maximise efficiency. In the next section we explore the effect Gd-loading concentration has on high energy neutron trigger efficiency.

\section{Dependence on Gadolinium Loading}
\label{sec:gadvar}
\begin{figure} [h]
    \centering
    \includegraphics[width=0.49\textwidth]{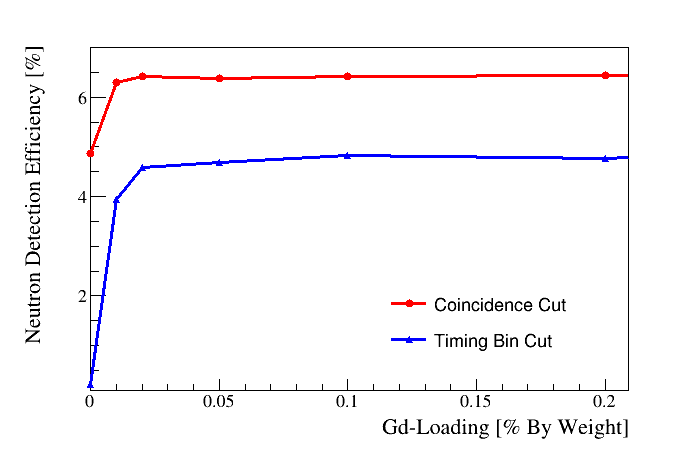}
    \includegraphics[width=0.49\textwidth]{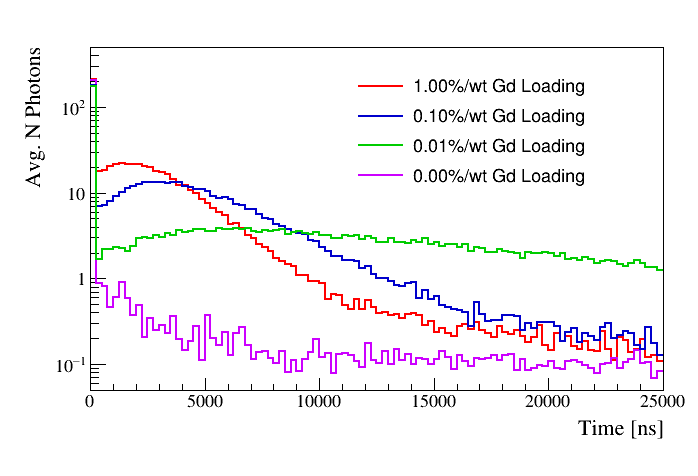}
    \caption{(Left) Efficiency of neutron detection plotted for varying gadolinium loading concentrations both before and after the number of flashes timing bin cut. (Right) Average number of photons in each timing bin distribution of neutrons after the number of flashes coincidence cut. The amplitude of each bin is significantly lower than the true number of photons observed in each event as the spallation event typically produces a small number of discrete flashes spread out in time.}
    \label{fig:gdloading}
\end{figure}

The cost of gadolinium is relatively low when compared to standard neutron detector technologies such as helium-3. However, when loading many water Cherenkov tanks to provide efficient coverage, the costs are still significant when compared to standard WCDs. Based on current prices of gadolinium sulphate and an assumed 0.2\% by weight loading, the cost to load the hypothetical detector system is approximately £2,527. Whilst 0.2\% loading has been found to be sufficient for efficient neutron tagging in experiments such as Super-Kamiokande the specific requirements for surface level Gd-WCDs are likely to differ as a result of differences in spallation neutron energies. To investigate the minimum Gd-loading required, simulations were repeated with only neutrons generated in CRY for varying gadolinium concentrations, and average efficiencies estimated for neutron energies between 10$^1$~MeV and 10$^4$~MeV. As shown in Figure \ref{fig:gdloading}, the efficiency of neutron detection when using the simplistic trigger described in the previous section plateaus at a loading fraction of 0.01-0.02\% by weight. This represents a saving of approximately £2,274 per detector. 

\begin{figure}
    \centering
    \includegraphics[width=0.49\textwidth]{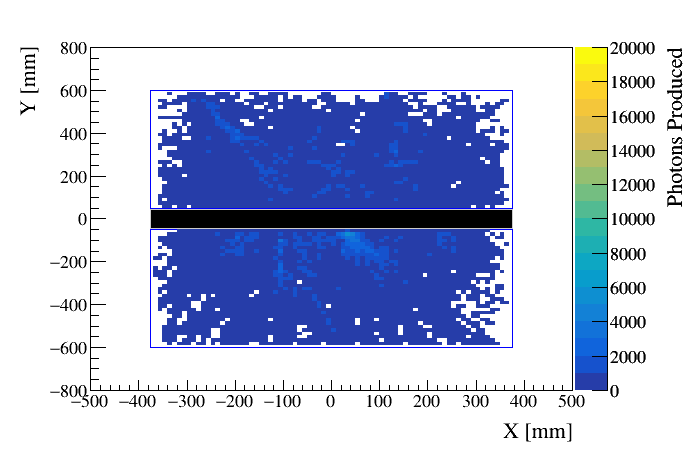}
    \includegraphics[width=0.49\textwidth]{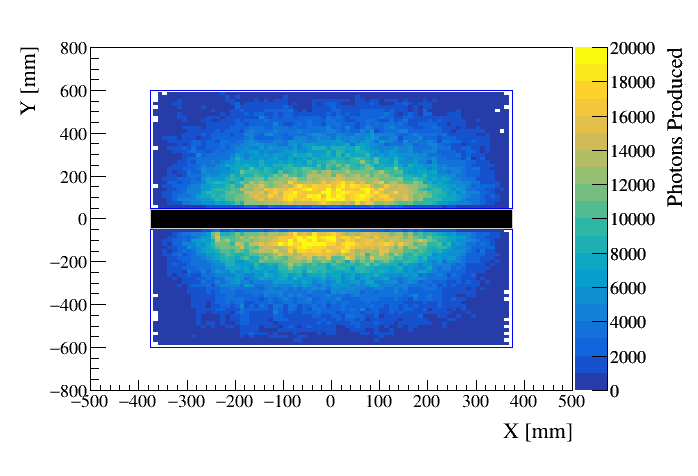}
    \caption{Distribution of Cherenkov photon product points for simulations of high energy neutron interactions. (Left) Standard unloaded WCD. (Right) Gd-loaded WCD.}
    \label{fig:pos}
\end{figure}

The spread of the timing distribution for neutron events was also found to increase with reductions in the Gd-loading concentration. This is a result of small increases in the capture cross-section of the Gd-Water mixture in each tank reducing the average path length of thermal spallation neutrons. Care must therefore be taken when optimising Gd loading concentration so as to not significantly extend the required deadtime of any detector system.

In addition to variations in the timing distribution, there was also a small change in the spread in positions of the Cherenkov photons produced in each tank, with increases in Gd concentration leading to more photons being generated closer to the central lead producer. Figure \ref{fig:pos} shows an extreme case of the difference between Cherenkov photon production points for high energy neutron interactions in normal WCD and Gd-WCD. It is clear that a large portion of the neutron captures occur within the first 10~cm either vertical side of the centre of the tank, suggesting that large area flat tank designs may offer an optimal  trade off between lead producer cost and gadolinium sulphate cost.

\section{Summary}
\label{sec:conc}

Through GEANT4 simulations of a hypothetical Gd-WCD tank design we have shown that it is possible to discriminate high energy cosmic ray neutrons from background particles using the timing signature of neutron induced spallation events. Importantly this signature is sensitive to the highest energy neutrons of an extended air shower, and can be optimised to have decay times that are comparable to the hold off of typical NM64 detectors. The advantage of this detector design is that it could be readily adopted by cosmic ray water Cherenkov detectors that use a split tank design (such as the upcoming SWGO). Given the clear timing signature of neutron spallation events it may also be possible to apply this technique inside single tank designs, however work is needed to understand whether event pile-up in ultra high energy showers could reduce overall detection sensitivity. 

In addition to the estimation of neutron content in high energy showers, the proposed detector design is also applicable to the long term monitoring of hazardous space weather. The high energy selection criteria effectively removes contributions from thermal neutrons which carry background systematic uncertainties from interactions with the soil surrounding a detector. It is therefore possible that this detection system could provide a lower cost alternative to the widely used NM64 neutron detector design. Based on the results shown in Figure \ref{fig:pos}, it is also likely that a low profile system could be developed that increases the overall surface area with a minimal increase in Gd loading cost by reducing the vertical detector height. This could be used to optimise the overall neutron counting rate for long term space weather monitoring, at the expense of degrading energy resolution for electromagnetic cosmic ray particles (gammas and electrons). 

\section*{Acknowledgements}
P.~Stowell would like to thank the Royal Commission for the Exhibition of 1851 for supporting this work. 

\bibliographystyle{unsrt} 
\bibliography{main}

\end{document}